\documentclass[graybox]{svmult}

\usepackage{mathptmx}       
\usepackage{helvet}         
\usepackage{courier}        
\usepackage{type1cm}        
\usepackage{makeidx}         
\usepackage{graphicx}        
\usepackage{multicol}        
\usepackage[bottom]{footmisc}

\makeindex             

\begin{document}

\title*{Tunnelling methods and Unruh-DeWitt detectors in curved spacetimes}
\author{Giovanni Acquaviva}
\institute{Giovanni Acquaviva \at in collaboration with Department of Physics, University of Trento, \email{gioacqua@gmail.com}}
%
%
\maketitle

\abstract{In this contribution we describe some interesting interplay between quantum theory, general relativity and thermodynamics.  In order to highlight the connection between these theories, we describe two approaches that allow to calculate thermal features as perceived by different observers in curved spacetimes. the tunnelling method and the Unruh-DeWitt detector.  In this context, the semi-classical tunnelling approach is applied to the issue of Hawking radiation and allows the calculation of the horizon temperature in a wide variety of scenarios.  The Unruh-DeWitt model is instead a quantum field-theoretical approach that should give a more exact answer in terms of transition rates between energy levels of an idealized detector.}

\section{Introduction}
\label{sec:1}
Since the theoretical discovery of thermal radiance from black holes made by Hawking \cite{hawk}, the connections between gravitational systems (GR), quantum theory (QT) and thermodynamics (TD) have become undeniable matter of interest, mainly because of the longstanding purpose of syncretizing GR and QT in a coherent way.  The problems that arise when trying to directly (and naively) quantize GR are just one indication that a new theory is desirable which includes both original theories in its very foundations.  Leaving aside the task of formulating such a theory (a task that the present work does not deal with), one is lead to consider instead scenarios in which semi-classical or even quantized fields are coupled to a classical geometric background.  This kind of approach could be regarded only as an \emph{effective} version of the full (yet hypotetical) quantum-gravitational theory, while at the same time it could be useful in order to highlight a first degree of interaction between GR and QT.

Many approaches have been brought forward in order to reproduce and extend Hawking's result, which essentially states that
\begin{equation}\label{full}
 \frac{\Gamma_{em}}{\Gamma_{abs}} = \E^{-\omega / T_H}\, ,
\end{equation}
where the probability for a quantum of energy $\omega$ to be emitted from a gravitational horizon is found to have a Boltzmannian form: hence the outgoing radiation can be expressed as a thermal state and the associated temperature is given by the surface gravity evaluated on the horizon, i.e. $T_H = \kappa_H/2\pi$.  One of the limitations of this result is the stationarity of the geometric background needed for its derivation.  

In this contribution we present two approaches that are able to confirm this result as well as to extend its range of validity to more general scenarios.  In Sect.~\ref{sec:2} we sketch the so-called tunnelling method: this is essentially a variant of the original method used by Hawking and - equipped with suitable ingredients - can be applied to dynamical (both black hole's and cosmological) horizons.  In Sect.~\ref{sec:3} instead a model of Unruh-DeWitt detector is presented: this approach makes use of quantum field-theoretical tools in order to build a \emph{quantum thermometer} endowed with a trajectory in a curved spacetime.  In Sect.~\ref{sec:4} we draw some conclusions.

\section{Tunnelling methods}
\label{sec:2}
It is well known that in a WKB approximation, a tunnelling probability rate is given by
\begin{equation}\label{rate}
 \Gamma \propto \E^{-2\, Im(S)}
\end{equation}
where $S$ is the classical action along the trajectory.  The presence of a non-vanishing imaginary contribution in $S$ is thus linked to a non-zero probability associated to the tunnelling trajectory.

\begin{figure}[h!!]
\includegraphics[width=110mm]{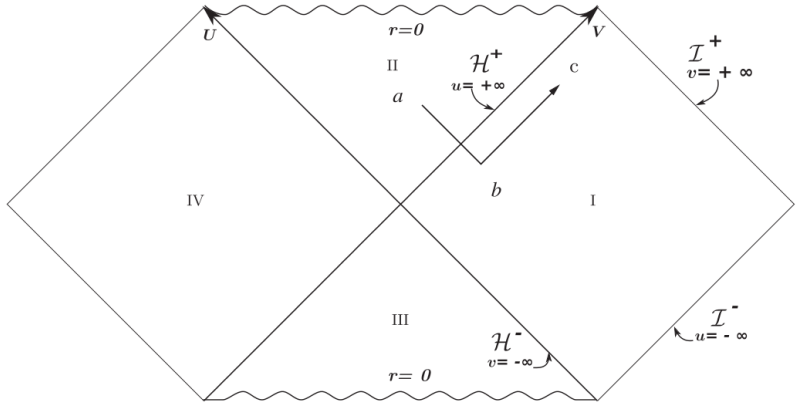}
\caption{Eternal Schwarzschild black hole.  The oriented null path $\vec{abc}$ crosses the horizon $\mathcal{H}^+$ once, before escaping towards null infinity.}
\label{fig:1}
\end{figure}

Historically one can actually identify two different (but equivalent in the stationary regime) approaches that fall under the denomination of tunnelling method:
\begin{itemize}
 \item the \emph{null-geodesic method}, introduced by Kraus, Parikh and Wilczek \cite{kraus, parikh} and
 \item the \emph{Hamilton-Jacobi method}, formulated by Padmanabhan and collaborators \cite{padma}
\end{itemize}
Here we will focus on the latter\footnote{See \cite{vanzo} for details of both methods and a discussion on why the H-J method is preferable.}.  The procedure to be followed is easily spelled in the following steps:
\begin{enumerate}
 \item the action $S$ of the massive tunnelling particle is assumed to satisfy the relativistic Hamilton-Jacobi equation
\begin{equation}\label{hj}
 g^{\mu\nu}\, \partial_{\mu}S\, \partial_{\nu}S\, +\, m^2\, =\, 0
\end{equation}
where $g^{\mu\nu}$ is the inverse metric of the spacetime considered;
 \item then one makes use of the identity
\begin{equation}\label{action}
 S\, =\, \int_{\gamma} \partial_{\mu}S\, \D x^{\mu}
\end{equation}
giving an ansatz for the form of the action based on the symmetries of the metric; the integration is carried along an oriented, piecewise null curve $\gamma$ crossing the horizon at least in one point, as shown in Fig.~\ref{fig:1} for an eternal black hole spacetime;
\item eventually one expresses the integrand of (\ref{action}) through (\ref{hj}) and performs a near-horizon approximation, treating the divergence through Feynman's $-\I\epsilon$ prescription\footnote{The choice of the sign in the prescription is related to the choice of positive-energy particles propagating towards infinity.}.
\end{enumerate}
The solution of the integral acquires a non-vanishing imaginary part,
\begin{equation}
 Im(S)\, =\, \frac{\pi\, \omega}{\kappa}
\end{equation}
where $\omega$ is the energy of the tunnelling particle and $\kappa$ is the surface gravity of the horizon.  Inserting the expression for the imaginary contribution in (\ref{rate}), and by comparison with the Boltzman distribution (\ref{full}), one can identify the temperature associated to the emitted radiation:
\begin{equation}
 T_H = \frac{\kappa}{2\pi}
\end{equation}
The result is easily extended to the whole Kerr-Newman class, for both scalar and fermionic tunnelling particles.

In order to treat the dynamical metric regime, the introduction of Kodama-Hayward theoretical results (see  \cite{kodama,hayward}) has been shown to be determinant in the spherically symmetric case.  Identifying the areal radius of the metric spheres with $R$, the definition of a Kodama vector is fairly simple: $K^{\mu} = \varepsilon^{\mu\nu}\partial_{\nu}R$.  This vector field can be regarded as a sort of generalization of what the Killing vector field in stationary cases is: in fact it allows to define a particle's conserved energy through its flow and a surface gravity associated to the \emph{trapping horizon}, respectively
\begin{eqnarray}
 \omega_H &&= -K^i\, \partial_i S\\
 \kappa_H &&= \frac{1}{2\sqrt{-\gamma}}\partial_i\left(\sqrt{-\gamma}\, \gamma^{ij}\partial_j\, R\right)\vert_H
\end{eqnarray}
where $\gamma_{ij}$ is the $(1+1)$ metric normal to the spheres of symmetry.  Applying the same procedure as before with these quantities in mind, the invariant imaginary contribution $Im(S) = \pi\omega_H/\kappa_H$ is found and one can thus identify the temperature $T_H = \kappa_H / 2\pi$ associated to the dynamical trapping horizon.

The main features regarding the tunnelling picture include:
\begin{itemize}
 \item the possibility of proving the covariance of the method as well as the equivalence of its two aforementioned versions;
 \item the consistency of the result in a wide variety of situations: higher-dimensional solutions, Taub and Taub-NUT solutions, decay of unstable particles, emission from cosmological horizons and naked singularities.
\end{itemize}

\section{Unruh-DeWitt detectors}
\label{sec:3}
As it is often reasonably argued, in the context of diffeomorphism invariant theories like GR one can not ignore the fact that \emph{particle} is an observer-dependent concept.  As a consequence, the description of phenomena like the Hawking effect should be better treated from the point of view of specific observers.  

The so-called Unruh-DeWitt detector \cite{unruh,dewitt} can in fact be used in order to distinguish whether an observer is in a vacuum or not by means of three ingredients: i) a real scalar field $\phi(x^0,\vec{x})$ coupled to ii) a curved metric $g_{\mu\nu}$ and iii) a localized two-level quantum system with energy eigenstates $\{ |E_0\rangle , |E_1\rangle \}$ and endowed with a trajectory $\left(x^0(\tau),\vec{x}(\tau)\right)$.  The idea is to evaluate the probability for the transition
\begin{equation}
 | 0 \rangle | E_0 \rangle (\tau_0) \quad \rightarrow\quad | \phi \rangle | E_1 \rangle (\tau_1)
\end{equation}
irrespective of the final state of the field $| \phi \rangle$, i.e. to witness whether or not the two-level system gets excited under the action of field modes while moving along a given trajectory in the curved metric.  The interpretation of such excitations in terms of particles is not straightforward: nevertheless, in certain cases, one can associate a thermodynamical meaning to the result.

Leaving aside the details of the derivation (see \cite{acqua}), the relevant quantity that allows such a calculation is the \emph{transition rate} given in general form by
\begin{equation}\label{rate1}
 \dot{F}(E) = 2\, \int_0^{\Delta\tau}\, Re\left[ \E^{-\I\, E\, s}\, W\left(x(\tau),x(\tau-s)\right) \right]\, \D s
\end{equation}
where $E$ is the energy gap of the detector, $\Delta\tau$ is the duration of the detection (the temporal window during which the coupling is switched on, from $\tau$ to $\tau'=\tau-s$) and $W$ is the Wightman function evaluated on the trajectory.  The particular form of the Wightman function clearly depends on the metric of the spacetime and on the trajectory of the detector.  Considering a conformally coupled scalar field in a 4-dimensional, conformally flat spacetime with conformal factor $a(x)$, it acquires the form
\begin{equation}\label{wight}
4\pi^2\,  W(\tau,s) = \frac{1}{\sigma^2(\tau,s)} = \frac{1}{a(\tau)a(\tau-s)\times \sigma_M^2(\tau,s)}
\end{equation}
where $\sigma^2$ (resp. $\sigma_M^2$) is the separation between $x^{\mu}(\tau)$ and $x^{\mu}(\tau')$ in the $g_{\mu\nu}$ metric (resp. in Minkowski spacetime).  The problem of the divergence of $W$ for $s\rightarrow0$ can be dealt with in a comfortable way thorugh a pole-subtraction scheme and the resulting expression of the transition rate for this case is
\begin{eqnarray}\label{rate2}
 \dot{F} = &&- \frac{E}{2\pi}\theta(-E)\ +\ \frac{1}{2\pi^2} \int_0^{\infty} \cos(E\, s)\left( \frac{1}{\sigma^2(\tau,s)} + \frac{1}{s^2} \right)\, \D s\ +\nonumber\\
 &&- \frac{1}{2\pi^2}\int_{\Delta\tau}^{\infty}\frac{\cos(E\, s)}{\sigma^2(\tau,s)}\, \D s
\end{eqnarray}
In this expression it is possible to identify three contributions: the first term regards the process of spontaneous emission; the second term is an asymptotic contribution, which evaluates the response of the detector during an infinite time of detection; the last term modifies the second one by taking into account the effects of a finite-time window of detection.\\

We now specialize to the case of two stationary spacetimes, de Sitter and Schwarzschild, considering a particular observer which is sitting at constant distance from the horizon (a so-called Kodama observer, in the terminology of \cite{acqua}).  In both cases the Wightman function has the form
\begin{equation}\label{stat}
 W(s) = \frac{1}{4\pi^2} \frac{\kappa^2}{4V\, \sinh^2\left( \frac{\kappa}{2\sqrt{V}}\, s \right)}
\end{equation}
where $\kappa$ is the horizon's surface gravity and $V$ is the Tolman redshift factor.  Equation (\ref{stat}) is i) stationary and ii) periodic in imaginary time, two conditions that qualify the Wightman function as \emph{thermal}.  This denomination is justified: the contribution of the asymptotic term -- calculated by summing the residues of the infinite poles in the lower half complex $s$-plane -- is given by
\begin{equation}
 \dot{F}_{\infty}(E) = \frac{1}{2\pi} \frac{E}{\exp\left(\frac{2\pi \sqrt{V}}{\kappa}\, E\right) -1}
\end{equation}
which is a Planckian distribution.  Hence the detector registers a thermal radiation at temperature $T = \frac{\kappa}{2\pi\sqrt{V}}$.  Some comments regarding this result are in order:
\begin{itemize}
 \item the redshift factor that appears in the expression of the temperature is consistent with Tolman's theorem on thermodynamical equilibrium in a gravitational field: the product $T\sqrt{-g_{00}}$ is constant -- in this case proportional to the horizon's surface gravity;
 \item thanks to the redshift factor, the expression for the temperature can be separated in two contributions $T^2=T_A^2 + T_H^2$: the first one is proportional to the acceleration of the detector and hence related to its proper motion (Unruh effect), while the second one is purely related to the presence of the horizon.
\end{itemize}
Eventually, one can also evaluate the contribution to the transition rate coming from the finite-time term in (\ref{rate2}).  The result is an oscillating behaviour exponentially damped in time.  No thermal contribution comes from this term, representing only a transient towards the equilibrium.

\section{Conclusions}
\label{sec:4}

Thet wo methods presented in this contribution have been widely adopted throughout the literature in order to shed some light on possible thermodynamical interpretations of the gravitational field (or specific observables thereof).  The results obtained in both frameworks are clearly consistent if one limits the analysis to stationary space-times.  Moreover, the main feature arising from the Unruh-DeWitt method is the explicit observer-dependence of the result (through the Tolman factor), to be compared with the observer-independence of the tunnelling's outcome.\\
It is in the non-stationary cases that the two methods show less consistency (see \cite{acqua}).  The roots of this deviation in more general cases could be due i) to an actually different interpretation of the results or perhaps ii) to the role of the observer, which has a crucial weight in the Unruh-DeWitt picture.  In the latter case, it should be possible to identify a particular observer whose measurement gives the tunnelling result as an outcome.

\begin{acknowledgement}
GA would like to thank Luciano Vanzo, Sergio Zerbini and Roberto Di Criscienzo for valuable discussions and groupwork that lead to the results presented in this contribution.
\end{acknowledgement}
%

%
%
%

\end{document}